\documentstyle[12pt]{article}
\def\PBHC{\hat{\bar{{\cal P}}}}
\def\PHC{\hat{{\cal P}}}
\def\hcO{\hat{{\cal O}}}
\def\hcOp{\hat{{\cal O}}^\prime}

\def\hcO{\hat{{\cal O}}}

\def\ih{(\imath\hbar)^{-1}}
\def\htO{\hat{\tilde{\Omega}}}
\def\htH{\hat{\tilde{H}}}
\def\hTh{\hat{\Theta}}
\def\hH{\hat{H}}
\def\hA{\hat{A}}
\def\hC{\hat{C}}
\def\hpi{\hat{\pi}}
\def\hl{\hat{\lambda}}
\def\hbC{\hat{\bar{C}}}

\def\hP{\hat{P}}
\def\hX{\hat{X}}
\def\hHp{\hat{H}^\prime}
\def\hXp{\hat{X}^\prime}

\def\hG{\hat{G}}
\def\hQ{\hat{Q}}
\def\hQp{\hat{Q}^\prime}
\def\htPs{\hat{\tilde{\Psi}}}
\def\hPs{\hat{\Psi}}
\def\ap{\alpha^\prime}
\def\gp{\gamma^\prime}
\def\bp{\beta^\prime}
\def\app{\alpha^{\prime\prime}}
\def\bpp{\beta^{\prime\prime}}
\def\NP#1#2{{\it Nucl.Phys}. {\bf B#1} {#2}}
\def\PR#1#2{{\it Phys. Rep.} {\bf #1} {#2}}
\def\PL#1#2{{\it Phys.Lett.} {\bf B#1} {#2}}
\def\CMP#1#2{{\it Commun. Math. Phys.} {\bf #1} {#2}}

\begin{document}

\thispagestyle{empty}

\baselineskip=0.6cm

\noindent P.~N.~Lebedev Institute Preprint     \hfill
FIAN/TD/12--93\\ I.~E.~Tamm Theory Department       \hfill
\begin{flushright}{August 1993}\end{flushright}

\begin{center}

\vspace{0.5in}

{\Large\bf ON THE EQUIVALENCE BETWEEN  }\\

\vspace{0.1in}

{\Large\bf THE UNIFIED AND STANDARD VERSIONS }\\

\vspace{0.1in}

{\Large\bf OF CONSTRAINT DYNAMICS }

\bigskip

\vspace{0.3in}
{\large  I.~A.~Batalin and I.~V.~Tyutin}\\
\medskip  {\it Department of Theoretical Physics} \\ {\it  P.~N.~Lebedev
Physical Institute} \\ {\it Leninsky prospect, 53, 117 924, Moscow,
Russia}$^{\dagger}$\\

\end{center}

\vspace{1.5cm}

\centerline{\bf ABSTRACT}

\begin{quotation}

The structure of physical operators and states  of the unified constraint
dynamics is studied. The genuine second--class constraints encoded are
shown to be the superselection operators. The unified constrained dynamics
is established to be physically--equivalent to the standard BFV--formalism
with constraints split.

\end{quotation}

\vfill

\noindent

$^{\dagger}$ E-mail address: tyutin@uspif.if.usp.br

\newpage

\setcounter{page}{2}

\newpage

\section{Introduction}

In previous papers [1-3] of the present authors a unified formalism has
been suggested for description of constrained dynamical systems without
making use of explicit splitting the constraints into the first-- and
second--class ones.

The generating equations, formulated in Ref.[1, 2], yield not only the
constraint algebra entirely, but also determine the fundamental phase
variable commutators to satisfy the Jacobi identity automatically. The
latter circumstance is quite nontrivial because of the classical limit
existence requirement.

In recent paper [3] the generating equations have been strengthened in
such a way that their solution under the corresponding boundary conditions
becomes determined uniquely up to a natural canonical arbitrariness.
Besides, the unified formulation of constrained dynamics was given there by
direct constructing  the Unitarizing Hamiltonian without making use of
splitting the constraints into the first-- and second--class ones.

The main purpose of the present paper is to show the formalism, developed
in Ref.[1-3], to be physically--equivalent to the standard
BFV--formalism [4-7] with second--class constraints presented explicitly
[8]. In fact, we confine ourselves by the quasiclassical approximation,
having in mind the standard reasoning for extension towards the exact
operator formulation. Besides, we suppose the degenerate metric, defining
the classical phase variable bracket, to be of a constraint rank.

The following remark seems to be relevant here. According to the Dirac
quantization scheme, the second--class constraint operators are considered
to equal to zero strongly. If one constructs the corresponding BRST--BFV
operator to be nilpotent, proceeding directly from the second--class
constraints as regarded to vanish strongly, then the nilpotent operator
constructed appears to have nontrivial unphysical cohomologies [9,10]. To
eliminate these cohomologies, it was suggested in paper [10] to introduce a
tower of ``ghosts for ghosts''. However, this scheme appears, in fact, to
imply the constraint splitting again.

In the scheme of Ref.[1-3], the Heisenberg equations of motion require for
the encoded second--class constraints to be conserved in time but not to
vanish strongly. This crucial circumstance allows one to eliminate the
nonphysical cohomologies.

The present consideration is necessarily brief. All statements are
formulated without proofs. The detailed consideration will be published
elsewhere.

\section{Outline of the unified formalism}

In this Section  we recall in brief the form of the unified constraint
algebra generating equations as well as the structure of the Unitarizing
Hamiltonian.

Let $\{\hat{\Gamma}^A\}$ be a set of the fundamental phase variable operators.
For the sake of simplicity we confine ourselves by the case in which all the
operators $\hat{\Gamma}^A$ are supposed to be Bosons.

Following the Refs.[1-3], we assign a classical ghost parameter $\Gamma^*_A$ to
each initial operator $\hat{\Gamma}^A$. Being the operators $\hat{\Gamma}^A$
bosonic, all the ghost parameters $\Gamma^*_A$ are Fermions.

Let $\hat{\Theta}^\alpha(\hat{\Gamma})$ be a total set of irreducible
constraint operators of the theory.
We assign a canonical pair of ghost operators $(\hat{C}_\alpha$,
$\PBHC{}^\alpha)$ to each constraint $\hat{\Theta}^\alpha(\hat{\Gamma})$.

So, we have the following list of values of the Grassmann parity
$(\varepsilon)$ and ghost number (gh) ascribed to the basic objects
introduced:

$$\varepsilon(\hat\Gamma^A)=0,\quad\varepsilon(\Gamma^*_A)=
\varepsilon(\hat{C}_\alpha)=\varepsilon(\PBHC{}^\alpha)=1, \eqno{(2.1)}$$

$$\hbox{gh}(\hat\Gamma^A)=0,\quad\hbox{gh}(\Gamma^*_A)=
\hbox{gh}(\hat{C}_\alpha)=-\hbox{gh}(\PBHC{}^\alpha)=1. \eqno{(2.2)}$$

Next, one introduces the generating operators $\hat{\Omega}$,
$\hat{\Delta}$, $\hat{\Omega}_\alpha$,

$$\varepsilon(\hat{\Omega})=\varepsilon(\hat{\Omega}_\alpha)=1,\quad
\varepsilon(\hat{\Delta})=0, \eqno{(2.3)}$$

$$\hbox{gh}(\hat{\Omega})=\hbox{gh}(\hat{\Omega}_\alpha)=1,\quad
\hbox{gh}(\hat{\Delta})=2, \eqno{(2.4)}$$
to satisfy the equations:

$$\ih[\hat{\Omega},\hat{\Omega}]=\hat{\Delta},\quad
\hat{\Delta}\left|_{{}_{\Gamma^*=0}}=0,\right. \eqno{(2.5)}$$

$$\ih[\hat{\Omega},\hat{\Omega}_\alpha]=0,\quad
\ih[\hat{\Omega}_\alpha,\hat{\Omega}_\beta]=0, \eqno{(2.6)}$$
together with their compatibility conditions:

$$\ih[\hat{\Omega},\hat{\Delta}]=0,\quad
\ih[\hat{\Delta},\hat{\Omega}_\alpha]=0. \eqno{(2.7)}$$

A solution to the generating equations (2.5), (2.6) is seeked in the form of
a series expansion in powers of the ghost parameters $\Gamma^*_A$ and
operators $(\hat{C}_\alpha$, $\PBHC{}^\alpha)$. The lowest  orders of these
expansions are:

$$\hat{\Omega}=\hat{C}_\alpha\hat{\Theta}{}^\alpha(\hat{\Gamma})+
\Gamma^*_A\hat{\Gamma}{}^A+\ldots, \eqno{(2.8)}$$

$$\hat{\Delta}=-2\hat{C}_\alpha\Gamma^*_A\hat{E}^{A\alpha}(\hat{\Gamma})
-\Gamma^*_B\Gamma^*_A\hat{D}^{AB}(\hat{\Gamma})+\ldots, \eqno{(2.9)}$$

$$\hat{\Omega}_\alpha=\hat{C}_\beta\hat{\Lambda}^\beta_\alpha(\hat{\Gamma})+
\Gamma^*_A\hat{K}^A_\alpha(\hat{\Gamma})+\ldots . \eqno{(2.10)}$$

The fundamental commutators:

$$\hat{\Sigma}^{AB}\equiv \ih[\hat\Gamma{}^A,
\hat\Gamma{}^B] \eqno{(2.11)}$$
are expressed in a certain way in terms of coefficient operators of the
expansions (2.8), (2.9).

As it has been shown in Ref.[3], a canonical transformation there exists such
that transformed operators $\hat{\tilde\Omega}$,
$\hat{\tilde\Delta}$, $\hat{\tilde\Omega}_\alpha$ are of the form:

$$\hat{\tilde\Omega}=\hat C_\alpha\hat{\tilde\Theta}{}^\alpha(\hat\Gamma)
+\Gamma^*_A\hat{\tilde\Gamma}{}^A(\hat{\Gamma}), \eqno{(2.12)}$$

$$\hat{\tilde\Delta}=-2\hat C_\alpha\Gamma^*_A
\ih[\hat{\tilde\Gamma}{}^A,\hat{\tilde\Theta}{}^\alpha]-
\Gamma^*_B\Gamma^*_A \ih[\hat{\tilde\Gamma}{}^A,
\hat{\tilde\Gamma}{}^B], \eqno{(2.13)}$$

$$\hat{\tilde\Omega}_\alpha=\hat{C}_\alpha, \eqno{(2.14)}$$
where $\hat{\tilde\Theta}{}^\alpha$ are Abelian constraints:

$$\ih[\hat{\tilde\Theta}{}^\alpha,\hat{\tilde\Theta}{}^\beta]=0.
\eqno{(2.15)}$$

To describe the unified dynamics, one introduces the new operator
$\hat\Phi$ to satisfy the equations:

$$\ih[\hat{\Phi},\hat{\Omega}_\alpha]=\hat{\Omega}_\alpha
\eqno{(2.16)}$$
and boundary condition:

$$\hat{\Phi}\bigl|_{{}_{\Gamma^*=0}}={1\over2}(\hat{C}_\alpha\PBHC{}^\alpha
-\PBHC{}^\alpha\hat{C}_\alpha)\equiv\hat G_0. \eqno{(2.17)}$$

Then, the truncated operator:

$$\hat{\Omega}_T\equiv\ih[\hat{\Phi},\hat{\Omega}],\quad
\varepsilon(\hat{\Omega}_T)=1,\quad\hbox{gh}(\hat{\Omega}_T)=1,
\eqno{(2.18)}$$
possesses the nilpotency property:

$$\ih[\hat{\Omega}_T,\hat{\Omega}_T]=0. \eqno{(2.19)}$$

In the tilded basis one has:

$$\hat{\tilde\Omega}_T=\hat C_\alpha\hat{\tilde\Theta}{}^\alpha.
\eqno{(2.20)}$$

Having the operator $\hat{\Omega}_T$ found, we define the Truncated
Hamiltonian $\hat{H}_T$,

$$\varepsilon(\hat{H}_T)=\hbox{gh}(\hat{H}_T)=0, \eqno{(2.21)}$$
to satisfy the equations:

$$\ih[\hat{H}_T,\hat{\Omega}_T]=0,\quad
\ih[\hat{\Phi},\hat{H}_T]=0 \eqno{(2.22)}$$
and boundary condition:

$$\hat{H}_T=\hat{H}_0+O(\PBHC), \eqno{(2.23)}$$
where $\hat{H}_0$ is the initial Hamiltonian of the theory. In the tilded
basis the operator $\hat{\tilde H}_T$ does not depend on $\Gamma^*_A$.

As a final step, we introduce canonical pairs of dynamically--active
Lagrangian multipliers $(\hat{\lambda}_\alpha$, $\hat{\pi}{}^\alpha)$ and
antighosts $(\PHC_\alpha$, $\hat{\bar C}{}^\alpha)$,

$$\varepsilon(\hat{\lambda}_\alpha)=\varepsilon(\hat{\pi}{}^\alpha)=0,\quad
\varepsilon(\PHC_\alpha)=\varepsilon(\hat{\bar C}{}^\alpha)=1, \eqno{(2.24)}$$

$$\hbox{gh}(\hat{\lambda}_\alpha)=\hbox{gh}(\hat{\pi}{}^\alpha)=0,\quad
\hbox{gh}(\PHC_\alpha)=-\hbox{gh}(\hat{\bar C}{}^\alpha)=1, \eqno{(2.25)}$$
to define the total BRST--BFV--operator:

$$\hat{Q}=\hat{\Omega}_T+\PHC_\alpha\hat{\pi}^\alpha \eqno{(2.26)}$$
and Unitarizing Hamiltonian:

$$\hat H=\hat H_T+\ih[\hat \Psi,\hat Q], \eqno{(2.27)}$$
where $\hat{\Psi}$ is a gauge Fermion operator,

$$\varepsilon(\hat\Psi)=1,\quad\hbox{gh}(\hat\Psi)=-1. \eqno{(2.28)}$$

Physical operators $\hat{\cal O}$ and physical states $|Phys\rangle$ are
defined in the usual way:

$$
\ih[\hcO,\hat Q]=0,\qquad
\ih[\hcO,\hG]=0,
\eqno{(2.29)}
$$

$$
\hQ|Phys\rangle =0,\qquad \hG|Phys\rangle=0,
\eqno{(2.30)}
$$
where $\hG$ is the total ghost number operator:

$$
\hG\equiv{1 \over 2}(\hC_\alpha\PBHC^\alpha-\PBHC^\alpha\hC_\alpha)+
{1 \over 2}(\PHC_\alpha\hbC^\alpha-\hbC^\alpha\PHC_\alpha)-
\imath\hbar\Gamma_A^*{\partial \over \partial \Gamma_A^*}
\eqno{(2.31)}
$$

It should be emphasized once again that we have used no constraint splitting
in constructing the Unitarizing Hamiltonian.

\section{The unified formalism is equivalent to the standard one}

In this Section we consider in details the structure of physical operators
and states. Then we establish the standard and unified versions of
constrained dynamics to be equivalent to each other.

First of all, being the fundamental bracket matrix
$\Sigma^{AB}\equiv\{\Gamma^A,\Gamma^B\}$ of a constant
rank, one can show that classical constraints are representable in the
following split form:

$$
\Theta^\alpha=M_{\bp}^\alpha\Theta^{\bp}+M_{\bpp}^\alpha\Theta^{\bpp}
\eqno{(3.1)}
$$
where $\beta=(\bp,\bpp)$,  $\bpp=1,\ldots ,$ corank
$||\Sigma^{AB}||$, the matrix
$||M_\beta^\alpha||=||M^\alpha_{\bp},M^\alpha_{\bpp}||$
is invertible, and the relations
hold:
$$
\{\Gamma^A,\Theta^{\app}\}=0,\qquad
\{\Theta^{\ap},\Theta^{\bp}\}=U_{\gp}^{\ap\bp}\Theta^{\gp}.
\eqno{(3.2)}
$$

Thus $\Theta^{\app}$ is nothing other but a set of genuine second--class
constraints, while $\Theta^{\ap}$ are first--class constraints.

It is quite natural that the constraint operators $\hat{\Theta}{}^\alpha$ of
the
quantum theory admit an ordered counterpart of splitting (3.1), so that one has
the corresponding operators $\hat{\Theta}^{\ap}$ and $\hat{\Theta}^{\app}$ to
be
quantum first-- and second--class constraints, respectively.

Now, let us turn to the tilded basis mentioned in the previous section. In
this basis the theory is determined by the equations:
$$
\ih[\htO_T,\htO_T]=0,\qquad
\ih[\htH_T,\htO_T]=0,
\eqno{(3.3)}
$$
where $\htO_T$  and $\htH_T$ do not depend on $\Gamma_A^*$. The same as in the
standard
case, one can confirm that the solution for $\htO_T$ is determined uniquely, up
to a canonical transformation, by the boundary value, while a change of the
constraint basis is also induced by a canonical transformation of $\htO_T$. The
solution for $\htH_T$ is determined, up to a contribution of the form
$\ih[\htPs,\htO]$, by
the boundary value, while a change $\hH_0\rightarrow\hH_0+O(\hTh)$ is
induced by a transformation of the form $\htH_T\rightarrow
\htH_T+\ih[\htPs,\htO_T]$.

Let us suppose that all the required canonical transformations are performed.
Besides, let us turn to the new phase variables,
$\hat\Gamma{}^A\rightarrow \hat{\Gamma}^{\prime A}=(\hat{\gamma}^i,
\hat\Theta^{\app})$, where the operators $\hat{\gamma}^i$, in principle, can be
considered to form a set of canonical pairs. Then in an appropriate basis the
operators $\hat{Q}$ and $\hat{H}$ possess the structure:

$$
\hQ=\hQp+\hat{Q}^{\prime\prime},\qquad
\hH=\hat{H}^{\prime}+\ih[\hPs,\hQ],
\eqno{(3.4)}
$$
where the nilpotent operator $\hQp$ is constructed in the usual
way, proceeding form the first--class constraints $\hTh^{\ap}$, and depends
on the variables $\hXp\equiv (\hat{\gamma}^i;
\hC_{\ap},\PBHC{}^{\ap},\PHC_{\ap},\hbC{}^{\ap},\hl_{\ap},\hpi^{\ap})$, while
the operator $\hQ^{\prime\prime}$ is of the form:

$$
\hQ^{\prime\prime}=\hC_{\app}\hTh^{\app}+\PHC_{\app}\hpi^{\app};
\eqno{(3.5)}
$$
$\hHp$ is the standard Unitarizing Hamiltonian of the BFV--formalism for the
variables $\hXp$, first--class constraints $\hTh^{\ap}$, nilpotent operator
$\hQp$
and boundary value
$$
H_0(\gamma)\equiv H_0|_{\Theta^{\prime\prime}=0}.
\eqno{(3.6)}
$$

Besides, the definition of the variables $\hX^\prime$ implies natural
splitting of the form:

$$ \hC_\alpha=(\hC_{\ap},\hC_{\app}),\ldots
\eqno{(3.7)}
$$
for the ghosts $\hC_\alpha$, $\PBHC^\alpha$ antighosts
$\PHC_\alpha$, $\hbC^\alpha$ and Lagrangian multipliers $\hl_\alpha$,
$\hpi^\alpha$.

Further, as the second--class constraints $\hTh^{\app}$ commute with all the
variables, the Heisenberg equations of motion, as applied to the
$\hTh^{\app}$, yield:
$$
d_t\hTh^{\app}=\ih[\hTh^{\app},\hH]=0, \eqno{(3.8)}
$$
and hence the conservation law holds:

$$
\hTh^{\app}=const.
\eqno{(3.9)}
$$
However, this constant is not to vanish certainly, but rather one should
considered $\hTh^{\app}$ to be the superselection operators [11]. The
mentioned circumstance appears to be of crucial importance for establishing
the srtucture of physical space.

Let us consider the structure of physical operators. We suppose the
operators $\hcO$ to be formal polynomials in all the variables. One
should consider all the operator equations, (2.29) for instance, to be
operator equalities valid for arbitrary values of $\hTh^{\app}$. Then one
can establish the physical operators $\hcO$ to have the structure:

$$
\hcO=\hcOp+\ih[\hA,\hQ],
\eqno{(3.10)}
$$
where $\hA$ is an arbitrary operator, while the operator $\hcOp$
depends on the variables $\hXp$ only, and satisfies the equation

$$
\ih[\hcOp,\hQp]=0. \eqno{(3.11)} $$
together with the condition

$$
\hbox{gh}(\hcOp)=0.
\eqno{(3.12)}
$$
Thus we conclude that the class of physical operators  $\hcO$ is equivalent
to the class of physical operators $\hcOp$ of the BFV--formalism for the
variables $\hXp$.

Let us consider the structure of physical subspace.  We suppose the total
space of states to be generated by applying the Heisenberg--field formal
polynomials to the cyclic vector (vacuum), the same as it holds in
theories without the superselection operators. This space is invariant
under the action of physical operators, the Hamiltonian or $S$-matrix for
instance. Besides, we suppose the vacuum $|0\rangle$ to be a physical
state.

Let:
$$
|Phys\rangle=\hP(\hX,\Gamma^*)|0\rangle,
\eqno{(3.13)}
$$
$$
\hQ|Phys\rangle=0,\qquad
\hG|Phys\rangle=0,
\eqno{(3.14)}
$$
where $\hX$ denotes the total set of variables of the extended phase space
operators, and $\hP$ is a formal polynomial.

It follows from  (3.13), (3.14) that:
$$
\ih[\hP,\hQ]=0,\qquad
\ih[\hP,\hG]=0.
\eqno{(3.15)}
$$
$i.e.$ $\hP$ is a physical operator.

Further, one can show that if the equations
$$
\hQ|Phys\rangle=0,\qquad
\hTh^{\app}|Phys\rangle=\xi^{\app}|Phys\rangle\neq 0
\eqno{(3.16)}
$$
hold, where $\xi^{\app}$ are some numbers, then

$$
|Phys\rangle=\hQ|\xi\rangle
\eqno{(3.17)}
$$
for some vector $|\xi\rangle$.

Thus the second--class constraints vanish effectively in the physical
subspace whose structure and dynamics coincide exactly with the ones of the
BFV--formalism for the basic operators $\hat{\gamma}^i$, first--class
constraints
$\hTh^{\ap}$ and Hamiltonian $\hH_0(\hat\gamma)$.

\section{Equivalence between the formalisms from the func\-tio\-nal--integral
viewpoint}

In this section we elucidate the equivalence between the unified and standard
versions of constrained dynamics from the functional--integral viewpoint.

Let us consider the generating functional of quantum Green's functions:

$$
Z(J)=\langle 0|T\exp\bigl\{\frac{\imath}{\hbar}
\int J_a\hX^a dt\bigr\}|0\rangle,
\eqno{(4.1)}
$$
where $J_a(t)$ is the external source, and

$$
\hX^a(t)\equiv(\hat{\Gamma}^A;\hC_\alpha,\PBHC^\alpha;
\PHC_\alpha,\hbC^\alpha;\hl_\alpha,\hpi^\alpha)
\eqno{(4.2)}
$$
are the Heisenberg operators,

$$
d_t\hX^a=\ih[\hX^a,\hH].
\eqno{(4.3)}
$$

As is usual, the following equations hold for the generating functional
(4.1):

$$
\biggl(\dot{X}^a-\{ X^a,H\}+J_b\{ X^a,X^b\}\biggr)|_{X=-\imath\hbar
\delta/\delta J}Z=0.  \eqno{(4.4)}
$$

Let us seek for a solution to these equations by making use of the
functional Fourier transform:

$$
Z(J)=\int DX\exp\bigl\{\frac{\imath}{\hbar}\int J_a
X^adt\bigr\}\tilde{Z}(X).
\eqno{(4.5)}
$$

It follows from (4.4) that:

$$
\tilde{Z}(X)=[\prod_t\prod_{\app}\delta(\dot{\Theta}^{\app}(t))]
\tilde{Z}_1(X),
\eqno{(4.6)}
$$
where $\Theta^{\app}$ are the second--class constraints, commuting
strongly with all the phase variables.

Thus the functional integral (4.5) appears to be concentrated on the
time--independent values

$$
\Theta^{\app}=const.
\eqno{(4.7)}
$$

At the present stage we have to formulate an important assumtion. Namely,
we suppose the representation (4.5) to be valid for the concrete
superselection sector. As for the functional integral itself, the above
assumption means that the integration is concentrated on such trajectories that
the functions $\Theta^{\app}(X(t))$ take asymptotically just the values
corresponding to the superselection sector chosen.

It has been established in the previous Section that the superselection
sectors with nonzero values of $\Theta^{\app}$ are physically--trivial.
The only nontrivial sector is the one with $\Theta^{\app}=0$. In this
sector one has:

$$
\prod_t\delta(\dot{\Theta}^{\app})=
(\hbox{Det}\frac{d}{dt})^{-1}\prod_t\delta(\Theta^{\app}).
\eqno{(4.8)}
$$

Performing a transformation to the representation, used in the previous
Section, and choosing in (3.4) a gauge Fermion\footnote{The same as in the
standard case, one can confirm that there exists a change of integration
variables that transforms the integrals with different
$\Psi$ into each other.} to be of the form

$$
\Psi=\bar{{\cal P}}^{\app}\lambda_{\app}
\eqno{(4.9)}
$$
one obtains:

$$
Z(J)=\int DX^\prime\exp\bigl\{\frac{\imath}{\hbar}\int[
\frac{1}{2}X^\prime(\{X^\prime,X^\prime\}^\prime)^{-1}\dot{X}^\prime-
H^\prime+J^\prime X^\prime]dt\bigr\},
\eqno{(4.10)}
$$
where external sources are introduced directly to the variables
$X^\prime$, the variables $\gamma^\prime$ are canonical, and
$\{\,,\,\}^\prime$ means the canonical brackets for all the variables
$X^\prime$

The expressions (4.10) coincides with the standard BFV--formalism
prescription.

{\bf Acknowledgement.} The authors are thankful to M.Henneaux for sending the
paper [10] together with interesting comments and to S.S.Horuzhy for
explaining some questions in the superselection operator theory.

\end{document}